\documentclass[aps,prb,amsmath,amssymb,reprint,superscriptaddress,preprintnumbers,showpacs,intlimits]{revtex4-1}
\usepackage{bm,latexsym,mathrsfs,enumerate,color}
\usepackage[mathcal]{euscript}
\usepackage[breaklinks=true,unicode=true,urlcolor = blue,colorlinks = true,citecolor = blue,linkcolor = blue]{hyperref}
\usepackage{graphicx}
\usepackage{pgf}
\renewcommand{\vec}[1]{\bm{#1}}
\usepackage[normalem]{ulem}
\begin{document}

\title{Stability of magnetic nanowires against spin-polarized current}

\author{Volodymyr P.~Kravchuk}
 \email{vkravchuk@bitp.kiev.ua}
 \affiliation{Bogolyubov Institute for Theoretical Physics, 03143 Kiev, Ukraine}

\date{\today}

%
%

\begin{abstract}
Stability of ground magnetization state of a thin magnetic nanowire against longitudinal spin-polarized current is studied theoretically with dipole-dipole interaction taken into account. The critical current, minimum current at which the instability of the ground state develops, is determined. Dependence of the critical current on size and form of the transversal wire cross-section is clarified. Theoretical predictions are confirmed by numerical micromagnetic simulations.
\end{abstract}


\maketitle
\section{Introduction}
Magnetic wires, whose transversal size is small enough to ensure the magnetization variation only along the wire, are of high applied interest now. These one dimensional magnetic systems are called nanowires an they are considered to be convenient elements for nonvolatile data storage devices of new type. \cite{Parkin08} Sequence of bits of information in such a wire is coded by sequence of magnetic domains magnetized along the wire. The magnetic domains are separated by domain walls of head-to-head and tail-to-tail configurations. \cite{Klaui08} Read-write processes require the motion of the domain sequence along the wire \cite{Parkin08}, this can be achieved by passing of pulses of spin-polarized current through the wire. \cite{Klaui05,Thiaville05,Parkin08} Recently it was shown \cite{Khvalkovskiy09} that one can significantly increase the domain wall velocity by applying the spin-polarized current perpendicularly to the wire. Since the usage of spin-polarized current is of high importance in this area, there arises a problem of stability of the uniform magnetization state against the current. Very recently we studied stability of the uniformly magnetized nanowires against perpendicular spin-polarized current. \cite{Kravchuk13} The stability analysis for the case of longitudinal current was considered somewhat earlier. \cite{Tserkovnyak08} However the dipole-dipole interaction was neglected in Ref.~\onlinecite{Tserkovnyak08}. In this paper we present a linear theory of stability of ground state of a long nanowire against the longitudinal spin-polarized current with the dipole-dipole interaction taken into account. In contrast to the previous results \cite{Tserkovnyak08} we show that due to the nonlocal nature of the dipole-dipole interaction the form and size of the wire transversal cross-section affect the stability condition. The analytical predictions are checked by numerical micromagnetic simulations.

\section{Model and linearized equation of motion}\label{sec:model}
Let us consider a rectilinear nanowire whose length $L$ much exceeds the characteristic transversal size. The wire is assumed to be narrow enough to ensure the magnetization uniformity in transversal direction. In the other words we assume that the magnetization varies only along the wire. The magnetic media is modeled as a discrete cubic lattice of magnetic moments $\vec M_{\vec \nu}$, where $\vec \nu=a(\nu_x,\nu_y,\nu_z)$ is a three dimensional index with $a$ being the lattice constant and $\nu_x,\nu_y,\nu_z\in\mathbb{Z}$. It is convenient to introduce
the following notations: $\mathcal{N}_z=L/a$ is the total number of lattice nodes along the $z$-axis oriented along the wire and $\mathcal{N}_s$ is the number
of nodes within the cross-section area.

The spin-polarized current with density $\vec j=j\hat{\vec z}$ is passed through the wire. Magnetization dynamics in this system is described by the modified Landau-Lifshitz-Gilbert equation \cite{Bazaliy98,Zhang04,Thiaville05} which can be written in the following discrete form
\begin{equation}\label{eq:LLG}
\begin{split}
&\dot{\vec m}_{n}=\left[\vec m_{n}\times\frac{\partial\mathcal{E}}{\partial\vec{m}_{n}}\right]+\alpha\left[\vec{m}_{n}\times\dot{\vec m}_{n}\right]\\
&-u\frac{\vec m_{n+a}-\vec m_{ n}}{a}+u\beta\frac{\vec m_{n}\times{\vec m_{n+a}}}{a}.
\end{split}
\end{equation}
Here the index $n=a\nu_z$ numerates the normalized magnetic moments $\vec{m}_n=\vec{M}_n/|\vec M_n|$ along the wire axis, the overdot indicates the derivative with respect to the dimensionless time measured in units $\omega_0^{-1}$ where $\omega_0=4\pi\gamma M_s$ with $\gamma$ being the gyromagnetic ratio, and $M_s$ being the saturation magnetization. $\mathcal{E}=E/(4\pi M_s^2a^3\mathcal{N}_s)$ is dimensionless total energy of the system. The normalized current is presented by the quantity $u=jP\hbar/(8\pi|e|M_s^2)$ which is close to average electron drift velocity, here $P$ is the rate o spin polarization, $\hbar$ is Planck constant and $e$ is electron charge. Here $\alpha$ is the Gilbert damping constant and $\beta$ is the nonadiabatic spin-transfer parameter.

First the problem of action of the spin-polarized conducting electrons on the magnetization states nonuniform along the current direction was discussed in Ref.~\onlinecite{Berger78}. The simple form of Eq.~\eqref{eq:LLG} without the nonadiabatic term was obtained in Ref.~\onlinecite{Bazaliy98} within the ballistic transport model for half-metallic materials. In this case the spin-wave instability of uniformly magnetized states was predicted for large currents. \cite{Bazaliy98,Li04} Later in Ref.~\onlinecite{Zhang04} the nonadiabatic spin-transfer term was introduced. The micromagnetic analysis of Eq.~\eqref{eq:LLG} was provided in Ref.~\onlinecite{Thiaville05} with corresponding study of the current driven domain wall motion. For detailed derivation of spin-torques and the applications see reviews \onlinecite{Ralph08a,Tserkovnyak08,Tatara08,Brataas12}.

In the following we use the previously developed method \cite{Gaididei12a,Kravchuk13} based on Holstein-Primakoff representation for spin operators \cite{Holstein40} generalized by Tyablikov. \cite{Tyablikov75} This method enables one to take into account the dipole-dipole interaction exactly for linear \cite{Volkov13d,Kravchuk13} as well as for weakly nonlinear \cite{Gaididei12a} problems. In line with the aforementioned method, we introduce the complex amplitude $\psi_{n}$ of the magnetization deviation from the ground state $\vec m=\hat{\vec z}$:
\begin{equation}\label{eq:psi}
\psi_{n}=\frac{m_{n}^x+i\,m_{n}^y}{\sqrt{1+m_{n}^z}},
\end{equation}
where $m_{n}^x$ and $m_{n}^y$ denote the magnetization components perpendicular to the wire.
In terms of the amplitude $\psi_{\vec n}$ the linearized form of \eqref{eq:LLG} reads
\begin{equation}\label{eq:Psi-lin}
  (1-i\alpha)\dot{\psi}_{n}=i\frac{\partial\mathcal{E}^0}{\partial\psi_{n}^*}-u(1-i\beta)\frac{\psi_{n+a}-\psi_{n}}{a},
\end{equation}
where $\mathcal{E}^0$ is harmonic part of the total energy\footnote{$\mathcal{E}^0$ includes terms not higher than $\mathcal{O}(|\psi_n|^2)$.}, for details see Appendix~\ref{app:LLG-exact}.

For the further analysis it is convenient to proceed to the wave-vector space, because in this case the energy $\mathcal{E}^0$ takes relatively simple form \cite{Gaididei12a,Kravchuk13} which enables us to proceed analytically. This is an advantage of the $\psi$-representation \eqref{eq:psi}. We use the one dimensional Fourier transform
\begin{subequations}\label{eq:Fourier-def}
\begin{align}
\label{eq:four-inv}&\psi_{n}=\frac{1}{\sqrt{\mathcal{N}_{z}}}\sum\limits_{k}\hat\psi_{k}e^{i k n},\\
\label{eq:four}&\hat\psi_{k}=\frac{1}{\sqrt{\mathcal{N}_{z}}}\sum\limits_{n}\psi_{n}e^{-i k n}
\end{align}
with the orthogonality condition
\begin{align} \label{eq:orth-cond}
\sum\limits_{n}e^{i( k- k')n}=\mathcal{N}_{z}\Delta(k-k'),
\end{align}
\end{subequations}
where $k=\frac{2\pi}{L}l$ is two-dimensional discrete wave vector, $l\in\mathbb{Z}$, and $\Delta(k)$ is the Kronecker delta. Applying \eqref{eq:Fourier-def} to the linearized equation \eqref{eq:Psi-lin} and using the long wave approximation $k\ll{2\pi}/{a}$ one obtains
\begin{equation}\label{eq:psi-four}
  (1-i\alpha)\dot{\hat\psi}_k=i\frac{\partial\mathcal{E}^0}{\partial\hat\psi_{k}^*}-uk(i+\beta)\hat\psi_{k}.
\end{equation}
\section{Energy of the system}
We consider here the case of a soft ferromagnet, therefore only two contributions into the total energy are taken into account: $E=E_{\mathrm{ex}}+E_{\mathrm{d}}$. Here
\begin{equation} \label{eq:Eex}
E_{\mathrm{ex}}=-\mathcal{S}^2\mathcal{J}\sum\limits_{\vec\nu,\vec\delta}\vec m_{\vec\nu}\cdot\vec m_{\vec\nu+\vec\delta}
\end{equation}
is the exchange contribution, where $\vec\delta$ numerates the nearest neighbors of an atom, $\mathcal{S}$ denotes value of the classical spin and $\mathcal{J}>0$ is exchange integral between two nearest atoms. In terms of the Fourier components $\hat\psi_k$ the harmonic part of the normalized exchange energy reads
\begin{equation}\label{eq:Eex-four}
\mathcal{E}_{\mathrm{ex}}^0=\ell^2\sum\limits_k k^2|\hat\psi_k|^2,
\end{equation}
where $\ell=\sqrt{\mathcal{S}^2\mathcal{J}/(2\pi M_s^2a)}$ is so called exchange length. The value of $\ell$ determines typical length-scale of the magnetization inhomogeneities, for typical magnets $\ell=2-10$ nm. \cite{Skomski03} The derivation of \eqref{eq:Eex-four} is analogous to one presented in Appendix A1 of Ref.~\onlinecite{Gaididei12a}.

The other term is the dipole-dipole energy
\begin{equation} \label{eq:Ems}
E_\mathrm{d}=\frac{M_s^2a^6}{2}\!\!\sum\limits_{\vec\nu\ne\vec\mu}\biggl[\frac{ (\vec m_{\vec\nu}\!\cdot\! \vec m_{\vec\mu})}{r_{\vec\nu\vec\mu}^3}
-3\frac{\left(\vec m_{\vec\nu}\!\cdot\! \vec r_{\vec\nu\vec\mu}\right) \left(\vec m_{\vec\mu}\! \cdot\! \vec r_{\vec\nu\vec\mu}\right)}{r_{\vec\nu\vec\mu}^5}\biggr],
\end{equation}
where we introduce the notation $\vec r_{\vec\nu\vec\mu}=(x_{\vec\nu\vec\mu},\,y_{\vec\nu\vec\mu},\,z_{\vec\nu\vec\mu})=\vec\mu-\vec\nu$.

Using that the magnetization depends only on $z$-coordinate one can write the harmonic part of the normalized dipole-dipole energy in form
\begin{subequations}\label{eq:Ed-harm}
\begin{align}\label{eq:Ed0-harm}
\mathcal{E}^0_\mathrm{d}=\frac12\sum\limits_k\bigl\{\left[\hat{g}(k)+2\hat{g}(0)\right]|\hat\psi_k|^2-3\hat{f}(k)\hat\psi_k\hat\psi_{-k}\bigr\}+c.c.,
\end{align}
see Appendix~\ref{app:Ed} for details. All information about form of the wire cross-section and its size is incorporated into functions
\begin{align}
&\hat{g}(k)=\frac{a^3}{8\pi\mathcal{N}_s}\sum\limits_n\sum\limits_{\begin{smallmatrix}\mu_x,\mu_y\\\nu_x,\nu_y\end{smallmatrix}}\frac{2n^2-x_{\vec\nu\vec\mu}^2-y_{\vec\nu\vec\mu}^2}{\left(x_{\vec\nu\vec\mu}^2+y_{\vec\nu\vec\mu}^2+n^2\right)^{5/2}}e^{ikn},\\
&\hat{f}(k)=\frac{a^3}{8\pi\mathcal{N}_s}\sum\limits_n\sum\limits_{\begin{smallmatrix}\mu_x,\mu_y\\\nu_x,\nu_y\end{smallmatrix}}\frac{(x_{\vec\nu\vec\mu}-iy_{\vec\nu\vec\mu})^2}{\left(x_{\vec\nu\vec\mu}^2+y_{\vec\nu\vec\mu}^2+n^2\right)^{5/2}}e^{ikn},
\end{align}
here we use the notation $x_{\vec\nu\vec\mu}=a(\mu_x-\nu_x)$ and $y_{\vec\nu\vec\mu}=a(\mu_y-\nu_y)$ for the sake of simplicity.
\end{subequations}

Let us consider a nanowire in form of tube with inner and outer radiuses $\rho$ and $R$ respectively. Applying the procedure of transition from the summation to integration with the singularity extraction (see Appendix B in Ref.~\onlinecite{Kravchuk13}) one obtains
\begin{equation}\label{eq:f-g-tube}
\begin{split}
&\left.\hat f(k)\right|_\mathrm{tube}=0,\\
&\left.\hat g(k)\right|_\mathrm{tube}=\frac{1}{R^2-\rho^2}\biggl[R^2\mathrm{I}_1(Rk)\mathrm{K}_1(Rk)-\\
-&2R\rho\mathrm{I}_1(\rho k)\mathrm{K}_1(Rk)+\rho^2\mathrm{I}_1(\rho k)\mathrm{K}_1(\rho k)\biggr]-\frac13,
\end{split}
\end{equation}
were $\mathrm{I}_1(x)$ and $\mathrm{K}_1(x)$ are modified Bessel functions of the first and second kind respectively \cite{NIST10}. In the limit case of cylindrical wire ($\rho\to0$) one obtains
\begin{equation}\label{eq:g2-cylinder}
  \left.\hat{g}(k)\right|_\mathrm{cyl}=\mathrm{I}_1(Rk)\mathrm{K}_1(Rk)-\frac13.
\end{equation}
Finally, the harmonic part of the dipole-dipole energy of cylindrical nanowire reads
\begin{equation}\label{eq:Ed-cyl}
  \left.\mathcal{E}^0_\mathrm{d}\right|_\mathrm{cyl}=\sum\limits_k\mathrm{I}_1(Rk)\mathrm{K}_1(Rk)|\hat\psi_k|^2.
\end{equation}
It should be noted that in case of nanowire with square cross-section the dipole-dipole energy has the similar form \cite{Kravchuk13}
\begin{equation}\label{eq:Ed-sqr}
  \left.\mathcal{E}^0_\mathrm{d}\right|_\mathrm{sqr}\approx\sum\limits_k\mathrm{I}_1\left({hk}/{\sqrt\pi}\right)\mathrm{K}_1\left({hk}/{\sqrt\pi}\right)|\hat\psi_k|^2,
\end{equation}
where $h$ is side of the square cross-section.
\subsection{Effective anisotropy approach}
Here we discuss a possibility to model the nanowire dipole-dipole energy by an easy axis anisotropy with the axis oriented along the wire:
\begin{equation}\label{eq:Ean}
E_\mathrm{an}=-\frac{K}{2}\sum\limits_{\vec\nu}(m^z_{\vec\nu})^2,\quad K>0.
\end{equation}
In the wave-vector space the harmonic part of normalized energy \eqref{eq:Ean} reads
\begin{equation}\label{eq:Ean-four}
  \mathcal{E}_\mathrm{an}^0=\kappa\sum\limits_k|\hat\psi_k|^2,
\end{equation}
where $\kappa=K/(4\pi M_s^2a^3)$. Comparing \eqref{eq:Ean-four} and \eqref{eq:Ed0-harm} one concludes that for a round nanowire the anisotropy constant is effectively $\kappa=\hat g(k)+2\hat g(0)$. Within the long-wave approximation $kR\ll1$, or in other words assuming that the characteristic size of magnetization nonuniformity much exceeds the transversal size of the wire, we obtain $\kappa\approx3\hat g(0)$. For the case of tubular or cylinder shaped nanowire the expression \eqref{eq:f-g-tube} results in the anisotropy constant $\kappa\approx1/2$. A few remarks should be made: (i) For the case of tubular wire with a thin wall ($\rho\approx R$) the simple form of anisotropy \eqref{eq:Ean} is insufficient, an additional easy-surface anisotropy term should be introduced. However this type of anisotropy can not be considered within the one-dimensional model which is used here and this discussion is beyond the scope of this paper. (ii) Accordingly to \eqref{eq:Ed-sqr} for a wire with square cross-section the effective anisotropy has the same value $\kappa\approx1/2$.

\section{Linear instability analysis}
Now we substitute into Eq.~\eqref{eq:psi-four} the energy expression $\mathcal{E}^0=\mathcal{E}^0_\mathrm{ex}+\mathcal{E}^0_\mathrm{d}$ where the exchange $\mathcal{E}^0_\mathrm{ex}$ and dipole-dipole $\mathcal{E}^0_\mathrm{d}$ contributions are determined by \eqref{eq:Eex-four} and \eqref{eq:Ed0-harm} respectively. Equation~\eqref{eq:psi-four} and its complex conjugated form compose a set of two linear equations with respect to functions $\hat{\psi}_k$ and $\hat{\psi}_{-k}^*$. The corresponding solutions are
\begin{subequations}
\begin{align}\label{eq:psi-sol}
  \hat{\psi}_k=\Psi_+e^{z_+(k)t},\quad \hat{\psi}_{-k}^*=\Psi_-e^{z_-(k)t}
\end{align}
where $\Psi_\pm$ are constants and the rate functions $z_\pm(k)$ are determined as
\begin{align}\label{eq:zpm-gen}
  (1+\alpha^2)z_\pm&=-\alpha\Omega-iuk(1+\alpha\beta)\pm\\\nonumber
&\pm\sqrt{\left[i\Omega+uk(\alpha-\beta)\right]^2+(1+\alpha^2)\varpi^2},
\end{align}
where we introduced the notations
\begin{align}\label{eq:omegas}
  &\Omega=\ell^2k^2+\hat{g}(k)+2\hat{g}(0),\\
&\varpi=\frac32\left|\hat{f}(k)+\hat{f}(-k)\right|.
\end{align}
\end{subequations}
Instability condition for the system can be written as
\begin{equation}\label{eq:inst-gen}
\exists k:\,\Re z_\pm(k)>0.
\end{equation}
In the following we consider the case $\varpi=0$, this corresponds to nanowires with symmetrical cross-sections: cylindrical rods, tubular and square nanowires. In this case the rate function has more simple form
\begin{subequations}
\begin{align}\label{eq:zpm-simple}
  z_\pm(k)=\frac{\gamma_\pm(k)\pm i\omega_\pm(k)}{1+\alpha^2},
\end{align}
where
\begin{align}\label{eq:gamma}
&\gamma_\pm(k)=-\alpha\left[\Omega(k)\pm uk\left(1-\frac{\beta}{\alpha}\right)\right],\\
\label{eq:omega}
&\omega_\pm(k)=\Omega(k)\mp uk(1+\alpha\beta).
\end{align}
\end{subequations}
The last summand in \eqref{eq:omega} represents the Doppler shift \cite{Fernandez-Rossier04,Tserkovnyak08} induced by the spin current.

The instability condition \eqref{eq:inst-gen} can be written now as $\gamma_\pm>0$ or equivalently
\begin{equation}\label{eq:inst-simple}
  |u|>u_c=\frac{U}{|1-\beta/\alpha|},\qquad U=\min\limits_{k>0}\frac{\Omega(k)}{k}
\end{equation}
The law $u_c\propto|1-\beta/\alpha|^{-1}$ was already obtained \cite{Tserkovnyak08} for anisotropic nanowires where the dipole-dipole contribution was neglected. In contrast to the previous results the expression \eqref{eq:inst-simple} takes into account form and transversal size of the wire which are incorporated into the shape parameter $U$.

\begin{figure}
\includegraphics[width=0.9\columnwidth]{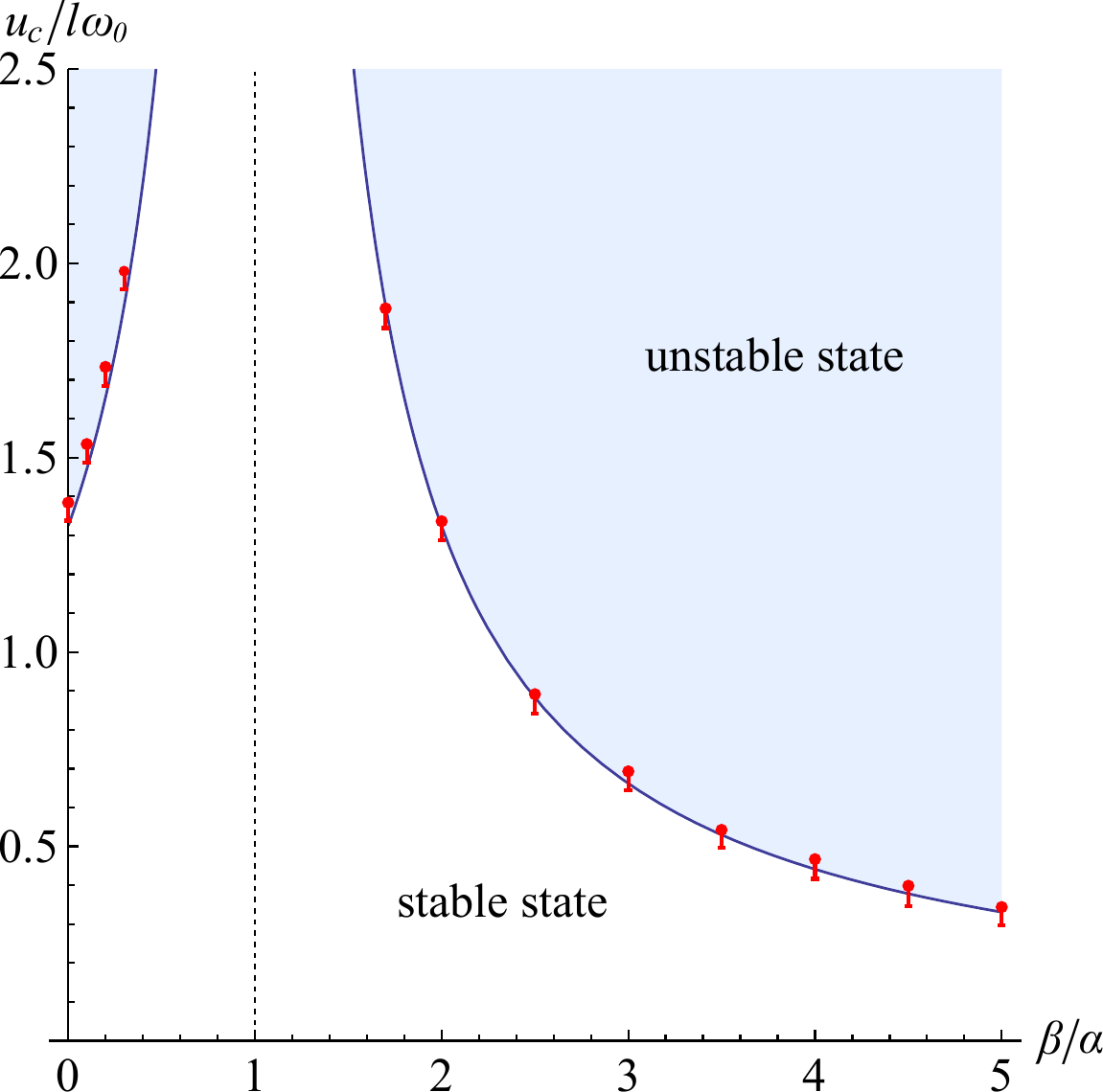}
\caption{The stability region for wire with square cross-section with $h/\ell=1.13$ (corresponds to $h=6$ nm for the case of permalloy). Solid line shows the critical current $u_c$ obtained from \eqref{eq:inst-simple} and \eqref{eq:Omega-sqr}. Transition to instability obtained with micromagnetic simulations is shown by verticals bars: in the top point and higher the instability is developed, in the bottom point and lower the state is stable.}\label{fig:stab-reg}
\end{figure}

As an example we consider a nanowire with square cross-section with side $h$. In this case
\begin{equation}\label{eq:Omega-sqr}
  \Omega(k)=\ell^2k^2+\mathrm{I}_1\left({hk}/{\sqrt\pi}\right)\mathrm{K}_1\left({hk}/{\sqrt\pi}\right)
\end{equation}
and the corresponding instability area determined by \eqref{eq:inst-simple} is shown in the Fig.~\ref{fig:stab-reg}.

\begin{figure}
\includegraphics[width=\columnwidth]{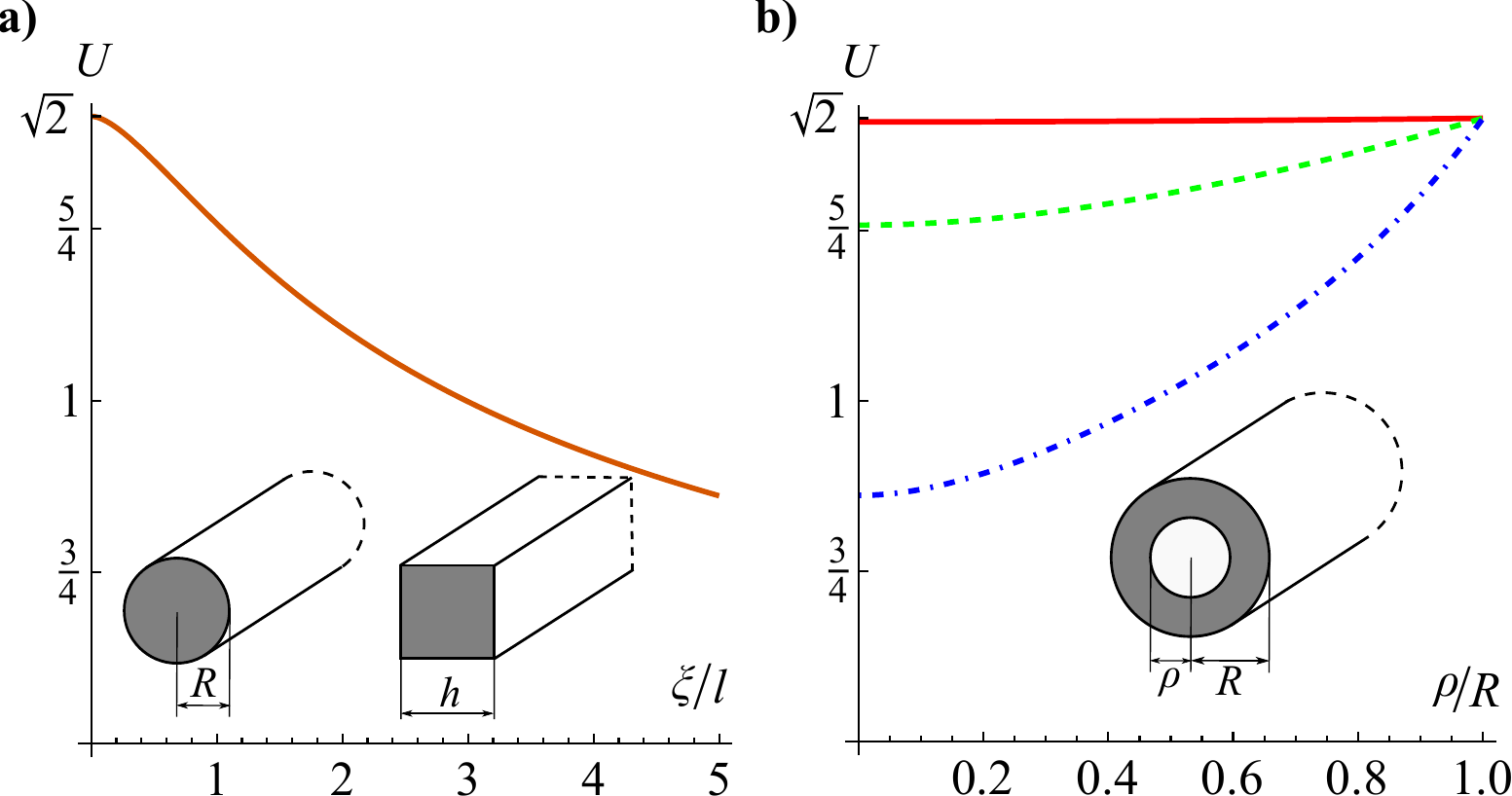}
\caption{The dependence of the shape parameter $U$ on form and size of the wire cross-section. Inset a) corresponds to the wires of round ($\xi=R$) and square ($\xi=h/\sqrt\pi$) cross-sections. Inset b) corresponds to the tubular wire with different outer radii: solid line -- $R/\ell=0.1$, dashed line -- $R/\ell=1$, dot-dashed line -- $R/\ell=5$.}\label{fig:U}
\end{figure}

The dependence of the shape parameter $U$ on form and size of the wire cross-section is demonstrated in the Fig.~\ref{fig:U}. As one can see the size and form dependence is noticeable.

To check the obtained stability condition \eqref{eq:inst-simple} we perform full scale micromagnetic simulations. \cite{OOMMF} We simulate the magnetization dynamics induced by the spin-current passing along a square nanowire with $h=6$ nm and $L=1\,\mu$m. The periodical boundary conditions are implemented along the wire. We choose material parameters of permalloy: saturation magnetization $M_s=8.6\times10^5$ A/m, exchange length $\ell=5.3$ nm (this corresponds to the exchange constant $A=1.3\times10^{-11}$ J/m). The anisotropy is neglected. The characteristic time scale is determined by the frequency of uniform ferromagnetic resonance $\omega_0=1.9\times10^{11}$~rad/s (30.3 GHz). The value of damping constant $\alpha=0.01$ is close to natural one. For permalloy the nonadiabatic spin-transfer parameter is $\beta=0.04$, \cite{Thiaville05} however we vary it in the range $0\le\beta/\alpha\le5$ in order to check the instability condition \eqref{eq:inst-simple}, see Fig.~\ref{fig:stab-reg}. The discretization mesh is cubic one: $\Delta x=\Delta y=\Delta x=3$ nm. The initial state is a slightly noised ground state $\vec m_\mathrm{ini}=\tilde{\vec m}/|\tilde{\vec m}|$, where $\tilde{\vec m}=(\tilde m_x,\tilde m_y,1)$ with transverse components $|\tilde m_x|<10^{-4}$ and $|\tilde m_y|<10^{-4}$ being determined in a random way. For a certain current value $u$ the magnetization dynamics is simulated during long time $\Delta t=100$ ns ($\sim 10^2\omega_0^{-1}\alpha^{-1}$). The judgement about stability is based on the time dependence of the total energy $E(t)$: if $E(t)$ exponentially decays then the ground state of the wire is considered to be stable for the given current $u$, and if the dependence $E(t)$ start to rice then the decision about instability is made. Results of the described stability analysis are shown in the Fig.~\ref{fig:stab-reg} by vertical bars: in the top point of the bar and higher the instability is developed, in the bottom point and lower the state is stable. One can see a nice agreement of the numerical results with the theoretical prediction \eqref{eq:inst-simple}.

In summary, we show that the dipole-dipole interaction noticeably changes the stability condition of the nanowire ground state with respect to the spin-current. Form and size of the wire cross-section affect the instability condition due to the nonlocal nature of the dipole-dipole interaction.

\section*{Acknowledgements}
The author is grateful to Prof. Yuri Gaididei and Prof. Denis Sheka for fruitful discussions. This work was supported by grant of NAS of Ukraine for young scientists (contract No. HM-85-2014).

\appendix
\section{Equation of motion in terms of amplitude $\psi$}\label{app:LLG-exact}
Considering $\vec m_n=\vec m_n(\psi,\psi^*)$ we project Eq.~\eqref{eq:LLG} to the transversal axes $x$ and $y$. Solving the obtained set of equations with respect to $\dot\psi$ and $\dot\psi^*$ one obtains
\begin{equation}\label{eq:psi-exact}
\begin{split}
  &(1+\alpha^2)\dot{\psi}_{n}=i\frac{\partial\mathcal{E}}{\partial\psi_{n}^*}\left(1+i\alpha\Psi_+\right)-\\
  &-u\frac{\psi_{n+a}-\psi_{n}}{a}\left[1+\alpha\beta+i(\alpha-\beta)\Psi_+\right]+\\
  &+\frac{\psi_{n}^2}{|\psi_{n}|^2}\Psi_-\left[\alpha\frac{\partial\mathcal{E}}{\partial\psi_{n}}+iu(\alpha-\beta)\frac{\psi_{n+a}^*-\psi_{n}^*}{a}\right],\\
  &\Psi_\pm=\frac12\left(\frac{2-|\psi_{n}|^2}{2}\pm\frac{2}{2-|\psi_{n}|^2}\right).
\end{split}
\end{equation}
For details see Appendix A of Ref.~\onlinecite{Kravchuk13}. Linearization of \eqref{eq:psi-exact} with respect to $\psi_n$ results in \eqref{eq:Psi-lin}.
\section{Dipole-dipole interaction for 1D case}\label{app:Ed}
As a direct consequence of dependence of magnetization on the longitudinal coordinate $z$ only the dipole-dipole energy \eqref{eq:Ems} can be presented in form
\begin{subequations}\label{eq:Ed-1D}
\begin{align}
&E_\mathrm{d}=\frac{M_s^2a^6}{2}\!\!\sum\limits_{\nu_z,\mu_z}\left[\sum\limits_{\varsigma=x,y,z}\mathcal{A}_{\nu_z\mu_z}^\varsigma m_{\nu_z}^\varsigma m_{\mu_z}^\varsigma+\mathcal{B}_{\nu_z\mu_z}m_{\nu_z}^x m_{\mu_z}^y\right],
\end{align}
where the summation over the transversal dimensions is enclosed in the coefficients
\begin{align}
&\mathcal{A}_{\nu_z\mu_z}^\varsigma=\sum\limits_{\begin{smallmatrix}\mu_x,\mu_y\\\nu_x,\nu_y\\\vec\nu\ne\vec\mu\end{smallmatrix}}\frac{r_{\vec\nu\vec\mu}^2-3\varsigma_{\vec\nu\vec\mu}^2}{r_{\vec\nu\vec\mu}^5},\quad\mathcal{B}_{\nu_x\mu_x}=-6\sum\limits_{\begin{smallmatrix}\mu_x,\mu_y\\\nu_x,\nu_y\\\vec\nu\ne\vec\mu\end{smallmatrix}}\frac{y_{\vec\nu\vec\mu}z_{\vec\nu\vec\mu}}{r_{\vec\nu\vec\mu}^5}.
\end{align}
\end{subequations}
Substituting now the magnetization components
\begin{equation}
\begin{split}
&m^z_n=1-|\psi_n|^2\\
&m_n^x\approx\frac{\psi_n+\psi_n^*}{\sqrt2},\quad m_n^y\approx\frac{\psi_n-\psi_n^*}{i\sqrt2}
\end{split}
\end{equation}
into \eqref{eq:Ed-1D} and applying the Fourier transform \eqref{eq:Fourier-def} one obtains harmonic part of the normalized dipole-dipole energy in form \eqref{eq:Ed-harm}.

%

\end{document}